\documentclass[a4paper,11pt]{article}
\pdfoutput=1 

\usepackage{jinstpub} 

\usepackage{etoolbox}

\usepackage [normalem]{ulem}
\graphicspath{{./figure/}}

\usepackage{caption}
\usepackage{booktabs}

\usepackage[table,dvipsnames,x11names]{xcolor}  
\usepackage[prefix=s]{xcolor-solarized}

\usepackage{soul}
\makeatletter
\patchcmd{\SOUL@ulunderline}{\dimen@}{\SOUL@dimen}{}{}
\patchcmd{\SOUL@ulunderline}{\dimen@}{\SOUL@dimen}{}{}
\patchcmd{\SOUL@ulunderline}{\dimen@}{\SOUL@dimen}{}{}
\newdimen\SOUL@dimen
\makeatother

\newcommand*\blue{\color{blue}}

\newcommand{\tblue}  [1]{\textcolor{blue}{#1}}

\renewcommand*\blue{\relax} \renewcommand*\tblue{\relax}

\newcommand*{\mltclmn}{\multicolumn}

\newcommand*{\dif}{\mathrm{d}} 

\usepackage{mathtools}

\usepackage{bm}

\begin{document}


\title{\boldmath The Magnet of the Scattering and Neutrino Detector for the SHiP experiment at CERN}


%
%

\collaboration{%
\includegraphics[height=17mm]{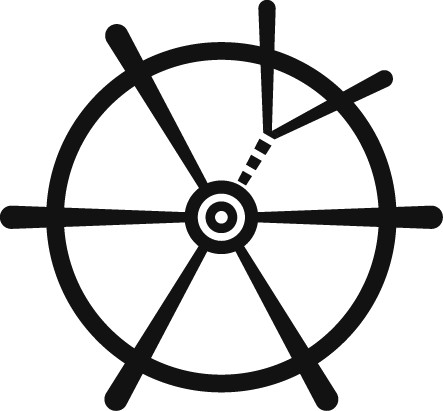}\\[6pt]
SHiP collaboration}

\author[44]{C.~Ahdida}
\author[14,d,h]{R.~Albanese}
\author[14]{A.~Alexandrov}
\author[39]{A.~Anokhina}
\author[18]{S.~Aoki}
\author[44]{G.~Arduini}
\author[38]{E.~Atkin}
\author[29]{N.~Azorskiy}
\author[54]{J.J.~Back}
\author[32]{A.~Bagulya}

% ...
\affiliation[14]{Sezione INFN di Napoli, Napoli, Italy}
% ...
\affiliation[18]{Kobe University, Kobe, Japan}
% ...
\affiliation[29]{Joint Institute for Nuclear Research (JINR), Dubna, Russia}
% ...
\affiliation[32]{P.N.~Lebedev Physical Institute (LPI), Moscow, Russia}
% ...
\affiliation[38]{National Research Nuclear University (MEPhI), Moscow, Russia}
\affiliation[39]{Skobeltsyn Institute of Nuclear Physics of Moscow State University (SINP MSU), Moscow, Russia}
% ...
\affiliation[44]{European Organization for Nuclear Research (CERN), Geneva, Switzerland}
% ...
\affiliation[54]{University of Warwick, Warwick, United Kingdom}
% ...
\affiliation[d]{Universit\`{a} di Napoli ``Federico II'', Napoli, Italy}
% ...
\affiliation[h]{Consorzio CREATE, Napoli, Italy}
% ...

\emailAdd{m.demagistris@unina.it, aquercia@unina.it}

\abstract{The Search for Hidden Particles (SHiP) experiment proposal at CERN demands
a dedicated dipole magnet for its scattering and neutrino detector. This requires
a very large volume to be uniformly magnetized at $B>1.2$\;T, with constraints regarding
the inner instrumented volume as well as the external region, where no massive structures are allowed
and only an extremely low stray field is admitted.
In this paper we report the main technical challenges 
and the relevant design options providing a comprehensive design for the magnet of the SHiP Scattering and Neutrino Detector.}

\keywords{Large detector systems for particle physics,
neutrino detectors, normal-conducting magnets.}

\arxivnumber{1910.02952} 




\maketitle
\flushbottom

\section{Introduction}
Given the absence of direct experimental evidence for Beyond the Standard Model (BSM)
physics at the high-energy frontier and the lack of unambiguous experimental hints for
the scale of new physics in precision measurements, it might well be that the shortcomings
of the Standard Model (SM) have their origin in new physics involving very
weakly interacting, relatively light particles. As a consequence of the extremely
feeble couplings and the typically long lifetimes,
{\blue the low mass scales for hidden particles are far less constrained~\cite{ship_th}.}
In several cases,
the present experimental and theoretical constraints from cosmology
and astrophysics indicate that a large fraction of the interesting
parameter space was beyond the reach of previous searches, but it is open
and accessible to current and future facilities. While the mass range
up to the kaon mass has been the subject of intensive searches,
the bounds on the interaction strength of long-lived particles above this
scale are significantly weaker. The recently proposed Search for Hidden
Particles (SHiP) beam-dump experiment \cite{ship} at the CERN Super Proton
Synchrotron (SPS) accelerator is designed to both search for decay signatures
by full reconstruction and particle identification of SM final states and
to search for scattering signatures of Light Dark Matter by the detection of recoil
\tblue{of atomic electrons or nuclei}.

The Beam Dump Facility (BDF) where SHiP operates
is well described
in ref.~\cite{ahdida2019}:
{the most upstream BDF part is a proton target followed by a 5\:m long hadron absorber}.
In addition to absorbing the hadrons and the electromagnetic
radiation, the iron of the hadron absorber is magnetised over a length of 4\:m.
Its dipole field makes up the first section of the
active muon shield~\cite{akmete2017}
which is optimised to sweep out of acceptance
{\blue muons of the entire momentum spectrum, up to 350\:GeV/$c$.}
The remaining part of the muon shield follows immediately
downstream of the hadron absorber in the experimental hall and consists of a
chain of iron core magnets which extends over a length of about 30\:m.

The SHiP experiment incorporates two complementary apparatuses.
The detector system immediately downstream of the muon shield is optimised
both for recoil signatures of hidden sector particle scattering and for neutrino physics.
It is based on a hybrid detector with a concept similar to what was developed by the
OPERA Collaboration~\cite{opera}
with alternating layers of nuclear emulsion films
with high-density $\nu$-target plates
and electronic trackers.
In addition, the detector is located in a magnetic field for charge sign and
momentum measurement of hadronic final states. The detector $\nu$-target mass totals
\tblue{about 10 tons}.
The emulsion spectrometer is followed by a muon identification system.
It also acts as a tagger for interactions in its passive layers which may produce
long-lived neutral mesons entering the downstream
decay volume and whose decay may mimic signal events.
The second detector system aims at measuring the visible decays of Hidden Sector
particles to both fully reconstructible final states and to partially
reconstructible final states with neutrinos. The detector consists of a 50\;m long
decay volume followed by a large spectrometer with a rectangular acceptance of 5\;m
in width and 10\;m in height~\cite{ahdida2019}.
The spectrometer is
designed to accurately reconstruct the decay vertex, the mass, and the impact
parameter of the hidden particle trajectory at the proton target.
A calorimeter and a muon identification system
provide particle identification.
A dedicated timing detector with $\sim$100 ps resolution provides a measure
of coincidence in order to reject combinatorial backgrounds. The decay volume
is surrounded by background taggers to identify neutrino and muon inelastic
scattering in the vacuum vessel walls.
The muon shield and the SHiP detector systems are
housed in a $\sim$120\:m long underground experimental hall at a depth of $\sim$15\:m.

In this paper we report the design and the expected performance of the
{\blue SND magnet, which contains the hybrid apparatus of the Scattering and Neutrino Detector}.
{\blue The work}
is organized as follows.
In section \ref{sec:exp:req} the experimental requirements at the basis of the overall design constraints are presented.
In section \ref{sec:mag:des} the full electromagnetic design is considered,
from analytical models to 3-D numerical simulations, defining a viable design option.
In section~\ref{sec:mech:issues}
the problem of mechanical forces and stresses are tackled,
along with functional issues relevant to the final mechanical structure of the SND magnet.
Finally section \ref{sec:concl} draws the conclusions.

\section{Experimental requirements\label{sec:exp:req}}
The design of the SHiP SND magnet follows the need for
a significantly large, uniformly magnetized volume, in order to accommodate
the $\nu$-target and the spectrometer trackers. This results in a magnetized
volume of about 10\:m$^3$ with a magnetic field of at least 1.2\:T.
The lower bound on the field strength comes from the requirement to measure
the charge sign and momentum of hadrons up to 10 GeV/$c$ in a very compact structure,
the so-called Compact Emulsion Spectrometer (CES)~\cite{fukushima2008},
made of 3 emulsion films interleaved with air over a total thickness of 3\:cm. 
At the same time, the stray field outside the magnet has to be sufficiently low (at the percent level
of the inner value) to avoid disturbing the flux of muons swept out by the muon shield.
This sets
{\blue severe constraints on the shape and size of the magnet.
In particular the magnet yoke, beside its fundamental
magnetic role (increasing efficiency,
{homogenizing and straightening up}
the field)
and the mechanical one (contrasting the strong magnetic expanding force acting on the coil),
is expected to sufficiently shield the field outside the magnet.
Such a requirement strongly affect the magnet design
{constraints and goals}.}
The detector mass and its operating temperature as well as the accessibility
for the detector installation and maintenance provide further challenges for the overall design.
In particular, the CES is supposed to be replaced every few weeks in order to
limit the total integrated flux of background muons, thus suppressing the
combinatorial background in the track matching required for the sagitta measurement.
{\blue That imply that the magnet has to be designed so that it can be frequently opened,
approximately once a fortnight.} 

{\blue
The required flux density over such a significant gap size requires a power of about 1\:MW.
In the past, at CERN, experimental magnets of comparable or even higher power consumption
(e.g.\ LHCb~~\cite{andre2002,andre2004,losasso2006} is 4.2\:MW)
were designed resistive to favour a much easier operation.
Furthermore, in this specific case, the CES will have to be replaced every few weeks and this will require
easy human accessibility, certainly more difficult in presence of helium and of a cryogenic infrastructure.
The resistive design reported here would however consume only one fourth of the LHCb magnet,
making the drawbacks of a superconducting version, including constructional difficulties, far more remarkable.
This is why for the baseline design we adopt a reliable and well-established design with resistive coils.
However, the study of a superconducting magnet will also be carried out, as an option to the baseline design.
One of the directions we intend to explore is an innovative concept of cryogen free magnet~\cite{tommasiniSupCu}
using HTS conductors, or alternatively LTS coils indirectly cooled with a small inventory of liquid helium.
This goes beyond the scope of this paper.
}

The power converter system and more generally any
ancillary
equipment
have to comply with CERN standard specifications.
Table~\ref{tab:magnet_spec_0}
reports the main specifications of the magnet.
%
\begin{table}[b]
\centering
\caption{\label{tab:magnet_spec_0}Magnet Specifications.}
\begin{tabular}{@{}lclc@{}}
\toprule
internal volume (detectors + ancillary equipment)     &     & [m$^3$] & $1\times1.6\times5.4$\\
overall external size                                 &     & [m$^3$] & $2.4\times4.0\times7.2$ \\
internal volume temperature                           &     & [$^{\circ}$C] & $18$\\
reference field (internal volume)                     & $B$ & [T] & $ > 1.2$\\
{\blue spatial field homogeneity} (internal volume)   & $\lvert \Delta B/B\rvert$ & [\%] & $\approx1$\\
{\blue temporal field stability (internal volume)}    & $\lvert \Delta B/B\rvert$ & [ppm] & $<10^3$\\
maximum external stray field                          & $B_\text{stray\,max}$     & [mT] & $\le 10$\\
\bottomrule
\end{tabular}
\end{table}

\section{Magnet design\label{sec:mag:des}}
\begin{figure}[!t]
\centering
\includegraphics[width=0.9\linewidth]{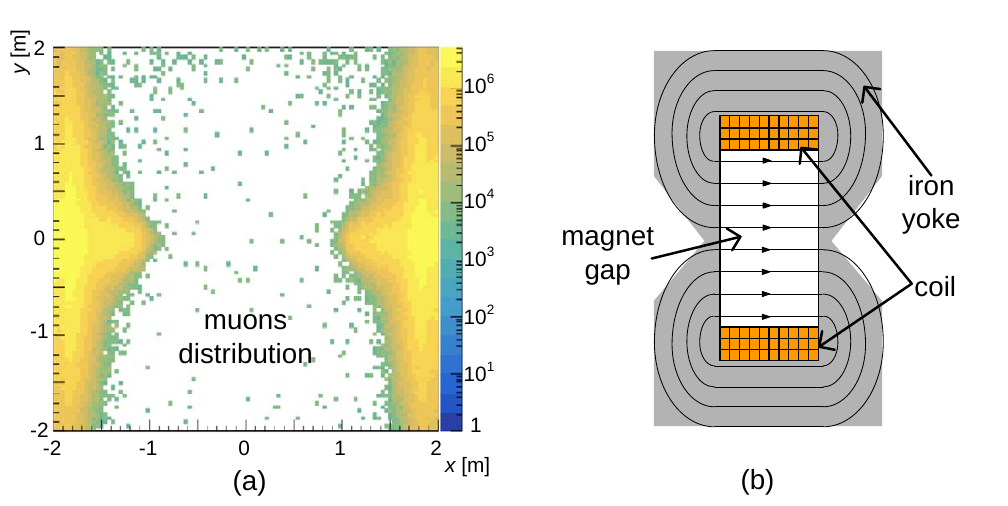}
\caption{\label{Fig_1}a) Simulated muons flux at the SND magnet position of the beam line.
  b) Sketch of the {\blue magnet's cross-section}.
  }
\end{figure}

\begin{figure}[!t]
\centering
\includegraphics[width=.999\linewidth]{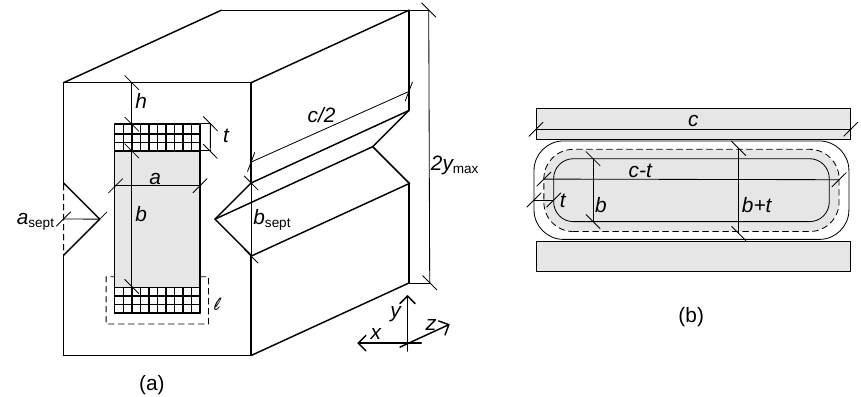}
\caption{\label{fig:Schematic_and_magnetic_circuit}
  {\blue (a) Schematic representation of one half of the magnet,
  showing the cross-section $z=0$ and the corresponding geometrical parameters.
  The point $x=y=z=0$ represents the centre of the magnet.
  The total magnet length along the direction of the beam (the $z$ direction) is equal to $c$.
  (b) Schematic $x=0$ cross-section, showing the areas (gray) considered for the flux balance.}
}
\end{figure}

{\mbox{Figure~\ref{Fig_1}a} shows the simulated profile of the muon flux distribution in the
transverse plane of the region where the SND detector is located.
\tblue{Such distribution}
sets
the fundamental constraint on the transverse shape of the magnet that does not have to
intercept the muon flux.} 
From this feature, the
{\blue magnet coil}
can be developed longitudinally, thus
providing a horizontal field and the inner magnetised volume can be taller than wider.
A conceptual design of the magnet is shown in figure~\ref{Fig_1}b
where the yoke shape is tapered according to the muon flux.  Figure~\ref{fig:Schematic_and_magnetic_circuit}
shows
a sketch
of the magnet structure, with the definition of major geometrical parameters.

\subsection{Analytical formulae \label{sec:an:mod}}
We describe now the procedure to get an approximate analytical magnetic model,
providing the basis for sizing the magnet.
The results of such analysis are then employed as the starting guess for the
detailed analysis that is performed in section~\ref{sec:int:mag:des},
including the electrical and thermal coil design.

{\blue
The standard design technique
which seeks the optimal current density, leading to total cost \mbox{minimization~\cite{green1967}},
cannot be adopted here.
In fact, we have a prescription on the maximum possible total magnet height,
which is fixed at $2y_\text{max}\cong4$\:m.
This is determined by the muons profile, which also sets the maximum tolerable external field
{$B_\text{stray\,max}$}
(\mbox{table~\ref{tab:magnet_spec_0}}).
The following analysis aims at determining design solutions that satisfy the
(internal and external) dimensional constraints
and the stray field specification,
while minimizing the power.
}

{\blue With reference to figure~\ref{fig:Schematic_and_magnetic_circuit}
we recognize the following fundamental geometric constraint
involving the coil and yoke thickness $t$ and $h$}
\begin{equation}
    \frac b2+t+h  =y_\text{max}
  \label{eq:geom_constraint}
\end{equation}
where $b=1.6$\:m is the total height
of the magnetized volume and $c=7.2$\:m is the magnet longitudinal length. 

{\blue By neglecting the stray flux, which is a reasonable assumption for a well designed yoke,
the flux is balanced when the internal flux $\phi_\text{int}$,
that is the sum of the fluxes corresponding to the gap and to the coil,
is equal to the flux into the yoke $\phi_\text{Fe}$.
That is easily done by considering the $x=0$ cross-section of the
magnet {(figure~\ref{fig:Schematic_and_magnetic_circuit}b)}.
The flux density in the coil decays approximately linearly, from the value $B$ at the internal edge
to zero at the outer edge. The flux in the coil, per unit length, is hence given by product $Bt/2$.
One then gets $\phi_\text{int} =(c-t)(b+t)B$,
where the product $(c-t)(b+t)$ is an average area
that takes into account the non uniformity of the flux density in the coil.
The flux balance equation $\phi_\text{int} =\phi_\text{Fe}$ then reads as
}
\begin{equation}
    (c-t)(b+t)B  =2hc {B_\text{Fe\,max}},
     \label{eq:flux_bal}
    \end{equation}
where $B_\text{Fe\,max}$ is the maximum value of the flux density, attained in the top
{\blue(and bottom)}
part of the yoke.

At this point we need to introduce the main figures of merit of the design, that are
magnet efficiency, electrical power, magneto-motive force and
stray field.

The magnet efficiency is defined as the ratio between the magnetic tension
over the gap and the magneto-motive force (MMF), or formally~\cite{tanabe2005,tommasini2011}
\begin{equation}
\eta ={\frac{aB/\mu_0}{NI}}
  =\frac{\int_\text{gap} \bm H\cdot \dif\bm\ell}{\int_\text{gap} \bm H\cdot \dif\bm\ell +\int_\text{iron} \bm H\cdot \dif\bm\ell}
    =\frac{1}{1+\frac1{\mu_r(B_\text{Fe\,max})} \frac{B_\text{Fe\,max}}{B}\frac\ell a}
  \label{eq:magn:effic:B}
\end{equation}
from which the following expression for the flux density $B$ is obtained
\begin{equation}
B  =\frac{\eta\mu_0NI}a  =\eta\mu_0ftJ,
\label{eq:transf:fun}
\end{equation}
being $N$ the number of coil turns, $I$ the current per turn and $J$ the current density,
$f=\frac{S_\text{active}}{at}$ the total filling factor,
$S_\text{active}$ being the area of the coil cross-section occupied by the conductor,
$\bm H$ the magnetic field
and $\mu_r$ the nonlinear yoke relative permeability.
Finally $\ell$ is the length of the line depicted in figure~\ref{fig:Schematic_and_magnetic_circuit}
corresponding to the region where $H$ is not negligible with respect to $H(B_\text{Fe\,max})$.
This, for low carbon steel yoke materials, yields $\ell \cong a + 2t$.

The above eq.~\eqref{eq:transf:fun}
allows to express both the MMF $\mathcal F$
and current density $J$ as a function of the flux density B. In particular, the former could be represented as 
\begin{equation}
  \mathcal F =NI =\frac{\mathcal F_\text{min}}\eta,
    \qquad  \mathcal F_\text{min} =\frac{aB}{\mu_0} 
    \label{eq:F:Fmin:eta}
\end{equation}
where $\mathcal F_\text{min}$ is the minimum value needed to get the expected $B$ ($\eta=1$).
From eq.~\eqref{eq:magn:effic:B} it is easy to realize that efficiency depends
on the effective magnet's working condition and, for a well-designed magnet, its values lie
in a range $\eta\approx0.95$ -- $0.98$~\cite{tommasini2011}. 

A key point is the estimation of the electrical power $P$ as a fundamental figure
of merit of the electromagnet,
which can easily be evaluated as follows.
{\blue The volumetric power density and the net volume occupied by the electrical conductor
are $\rho J^2$ and $\varOmega=fatl_t$, respectively,
where $\rho$ is the electrical resistivity of the conductor and $l_t\approx2(b+c)$ is the mean turns length.
Then, by using eq.~\eqref{eq:transf:fun}
one gets}
\begin{equation}
P =\int\limits_{\varOmega} \rho J^2\, \dif\varOmega
  \;=\frac{\rho }{\eta^2\mu_0^2}\frac{al_t}{ft}B^2
  \;\approx \frac{2\rho}{{\eta ^2 \mu _0^2 }}\frac{{a\left( {b + c} \right)}}{{ft}}B^2\,.
\label{eq:P:params:B2}
\end{equation}

The maximum stray field value, attained at the surface of the yoke,
is estimated by applying the continuity of the tangential component of $\bm H$
at the symmetry point $x=0$, $y=y_\text{max}$, $z=0$,
which reads as
\begin{equation}
B_\text{stray\,{max}} =\mu_0H_\text{stray\,{max}} =\mu_0H_\text{Fe\,max}
  = \frac{B_\text{Fe\,max}}{\mu_r(B_\text{Fe\,max})} =\frac{1-\eta}\eta \frac a\ell B
\label{eq:Bstray:mur:eta:a:B}
\end{equation}
{\blue where the rightmost equality follows from eq.~\eqref{eq:magn:effic:B}.}

\begin{figure}[tbp]
\centering
\includegraphics[width=.64\linewidth]{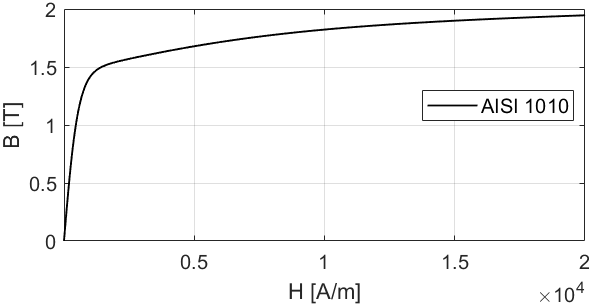}
\caption{\label{fig:AISI1010:comsol}
 Reference AISI 1010 $H$-$B$ curve.}
\end{figure}

Having defined the above quantities, the task is now the estimation of iron and corresponding coil thickness
such that the geometric and physical constraints specified in table~\ref{tab:magnet_spec_0}, are fulfilled,
after a certain choice for the iron material is made.
As basic reference we consider a typical AISI\:1010 $H$-$B$ curve, shown in figure~\ref{fig:AISI1010:comsol}.
 
%
\begin{figure}[tbp]
\centering
\includegraphics[width=.85\linewidth]{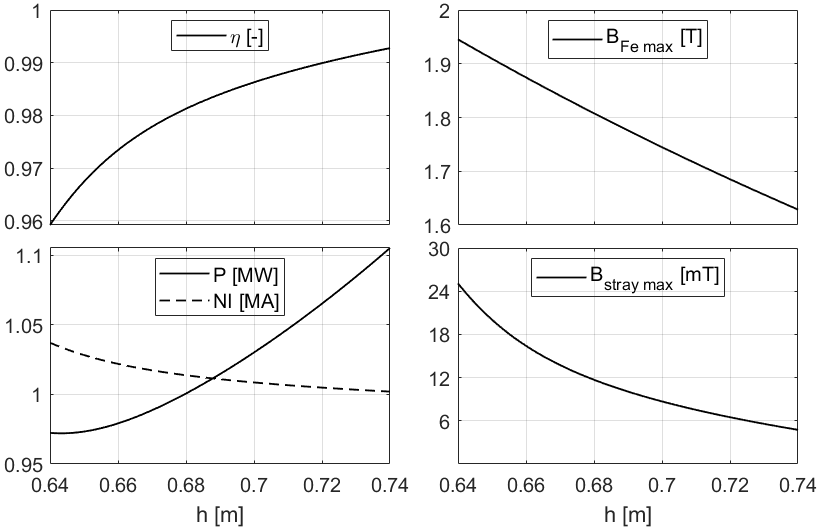}
\caption{\label{fig:an:mod:results}
 Dependence of efficiency, maximum iron flux density, power, total current and
 maximum stray flux density, as a function of the yoke thickness $h$, as predicted by the
 model~\eqref{eq:geom_constraint}--\eqref{eq:magn:effic:B}, \eqref{eq:F:Fmin:eta}--\eqref{eq:Bstray:mur:eta:a:B}.}
\end{figure}

By solving eqs.~\eqref{eq:geom_constraint}--\eqref{eq:flux_bal}
while assuming $h$ as parameter, one gets $t=y_\text{max}-b/2-h$,
$B_\text{Fe\,max}=(c-t)(b+t)B/(2hc)$.
In turn, eqs.~\eqref{eq:magn:effic:B}, \eqref{eq:F:Fmin:eta}--\eqref{eq:Bstray:mur:eta:a:B}
and the mentioned $H$-$B$ curve
we get $\eta$, $NI$, $P$ and $B_\text{stray\,{max}}$.
The analysis has been carried out by assuming $B=1.25$\:T, the geometrical parameters
as described above, leading to the plots shown in figure~\ref{fig:an:mod:results}.

From the inspection of the curves
it is easily realized that the stray field decreases with increasing $h$.
Conversely, the MMF shows weak variations with $h$
{\blue and tends toward its limit value (about 1\:MA, eq.~\eqref{eq:F:Fmin:eta}).}
The power is evaluated by assuming the following values for copper
resistivity $\rho=1.87\cdot10^{-8}\:\Omega$\,m (@\;$T=42.5^\circ$C), and the filling factor,
$f=0.65$,
which is a reasonable value in the coil design.
The corresponding curve shows a minimum for a specific value of the iron thickness,
which provides a significant information to be exploited for the design.
Minimising the stray field and the power at the same time results as conflicting goals.

{\blue Notice that one normally expects a completely different behaviour of the power,
namely a reduction of $P$ when $NI$ gets reduced.
In our case we have $P\propto1/(\eta^2t)$,
which is the product of a decreasing ($1/\eta^2$) and of an increasing ($1/t$) function of $h$, respectively.
The result is a power function that has a minimum and then increases with $h$, instead of decreasing.
This is due to the dimensional constraint~\eqref{eq:geom_constraint},
a specific peculiarity of the present design.}

Within the considered model the $B_\text{stray}\le 10\;$mT constraint is
achieved with $h = 0.7\;$m ($B_\text{stray} = 8.7\;$mT), which yields to $\eta = 0.986$,
{$P = 1.03\;$MW},
$\mathcal{F} = 1.01\;$MA and $B_\text{Fe\,max} = 1.74\;$T.
However, it should be outlined that such quantities are highly sensitive
to parameters variations.
In particular, from eqs.~\eqref{eq:flux_bal},~\eqref{eq:Bstray:mur:eta:a:B}
the stray field sensitivity to variation of the yoke thickness $h$
can be expressed as
\begin{equation}
  \frac{\dif B_\text{stray\,max}/B_\text{stray\,max}}{\dif h/h}
    =-\frac{c'b'+h^2}{(c' +h)(b' -h)} \frac{\mu_\text{r}(B_\text{Fe\,max})}{\mu_\text{r\,diff}(B_\text{Fe\,max})}
    \approx -10
  \label{eq:an:mod:stray:h:sens}
\end{equation}
where $c'=c+\tfrac b2 -y_\text{max} =6\:\text m$, $b'=\tfrac b2 +y_\text{max} =2.8\:\text m$,
$\mu_\text{r\,diff}$ is the differential relative permeability
and $h\approx0.7$.
%
%
The sensitivity resulting by considering different $H$-$B$ curves will be shown in next sections.

Finally it should be outlined that both copper and aluminum~\cite{green1967}
were considered as
materials for the conductor coil. However, since the limit in the coil size $t$
given by equation \ref{eq:geom_constraint}, the higher electrical resistivity
of aluminium
and the linear dependence of the power on $\rho$, a much larger dissipated
power results from aluminum choice, which is not compatible with current CERN standards.
For this reason, all the discussions carried out hereafter will refer to copper coils.

\subsection{Integrated magnet design\label{sec:int:mag:des}}
This section describes the magnet design with due detail. It includes yoke, coil and thermal shield detailed design,
with the main goal of trying to keep the ohmic power $P$ as low as possible while taking into account geometrical,
electrical and thermal aspects. In particular: i) the yoke design accounting for specific iron magnetic properties;
ii) the full electrical and cooling coil design, with constraints from integration of power supply according to
CERN standards and iii) the integrated design of the thermal conditioning achieving the required temperature
of the detector region. The due verification of the compatibility of such design option with mechanical loads,
forces and stresses is treated in the following section \ref{sec:mech:issues}.

\subsubsection{Yoke}
As stated in previous section, {three} geometrical parameters are specified by the design constraints, namely:
\begin{itemize}
\item {total} longitudinal magnet length $c = 7.20$ m;
\item horizontal gap $a = 1.00$ m
\item total height $2y_\text{max} = 4.00$ m
\end{itemize}

The simplified analytical model shows (figure~\ref{fig:an:mod:results})
that the limiting factor is the requirement to keep the stray field outside
the magnet
within the threshold $B_\text{stray} \le 10$\:mT specified in
section~\ref{sec:exp:req},
yielding a yoke thickness $h=0.70$\:m.

Moreover, this choice for the yoke thickness
provides a good magnet efficiency
$\eta=0.986$ and a power $P=1.03$\:MW, which is not far from the unconstrained minimum of $P=0.97$\:MW.
As for the geometrical parameters of the triangular septum $b_\text{sept}$ and $a_\text{sept}$ illustrated in
figure~\ref{fig:Schematic_and_magnetic_circuit}
they are selected so as to minimize the interaction with the muons,
whose distribution is depicted in figure~\ref{Fig_1}a, as triangular septum height $b_\text{sept}=1.44$\:m,
and triangular septum width $a_\text{sept}=0.50$\:m.

Some further considerations are due in terms of yoke material properties. Among yoke material types used at CERN
there are, ordered by performances (and cost), low carbon steels, such as AISI~1010,
special grade low carbon steels of relatively high purity, such as ARMCO$^\text{\textregistered}$ grade~4,
and cobalt iron.
Ref.~\cite{sgobba2011}
reports $H$-$B$ curves of materials used as magnetic steel as they are
obtained from measured samples,
in particular different heats of 1010 steel and a special grade one. We consider their upper and lower bounds,
that is the curves having the largest
and lower $B$ strengths,
which are labelled ARMCO ATLAS and ST\:1010 ATLAS in ref.~\cite{sgobba2011}.
In the region of interest, corresponding to $B_\text{Fe\,max}=1.75$\:T,
the working point is identified by $H(B_\text{Fe\,max})=6.02$\:kA/m and $H(B_\text{Fe\,max})=9.15$\:kA/m
for the two bounding materials, respectively.

\begin{figure}[tbp]
\centering
\includegraphics[width=.99\linewidth]{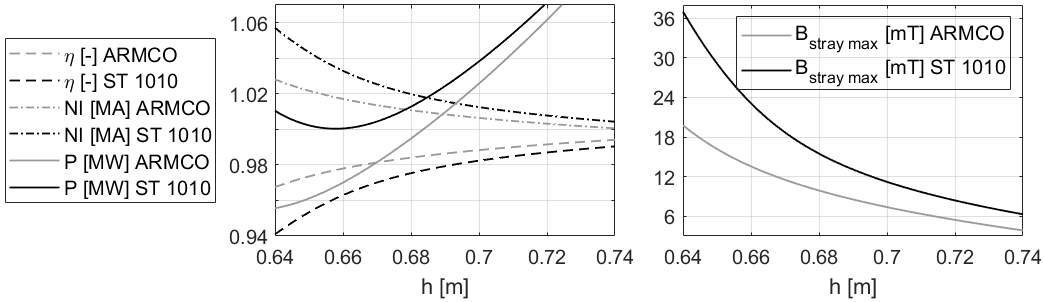}
\caption{\label{fig:different:iron}
 Dependence of efficiency, total current, power and
 maximum stray flux density, as a function of the yoke thickness $h$, as predicted by the
 model~\eqref{eq:geom_constraint}--\eqref{eq:magn:effic:B}, \eqref{eq:F:Fmin:eta}--\eqref{eq:Bstray:mur:eta:a:B},
 for yoke material curves corresponding to the considered upper and lower bounds, see text.}
\end{figure}

Figure~\ref{fig:different:iron}
shows the results obtained with the
model~\eqref{eq:geom_constraint}--\eqref{eq:magn:effic:B}, \eqref{eq:F:Fmin:eta}--\eqref{eq:Bstray:mur:eta:a:B}
and such bounds for iron characteristics.
There is a clear evidence of the sensitivity with respect to the material.
In particular, the variation of $P$ and $B_\text{stray\,max}$ is relevant in the region of interest. 
From now on we  assume the ST1010 ATLAS as material for the yoke, being the most conservative choice.
It is clear there is room for improvement by using better materials.

\subsubsection{Coil\label{sec:imd:coil}}
\begin{figure} 
\centering
\begin{tabular}{@{}c@{\qquad}c@{}}
  \multicolumn2c{\includegraphics[height=5.50cm]{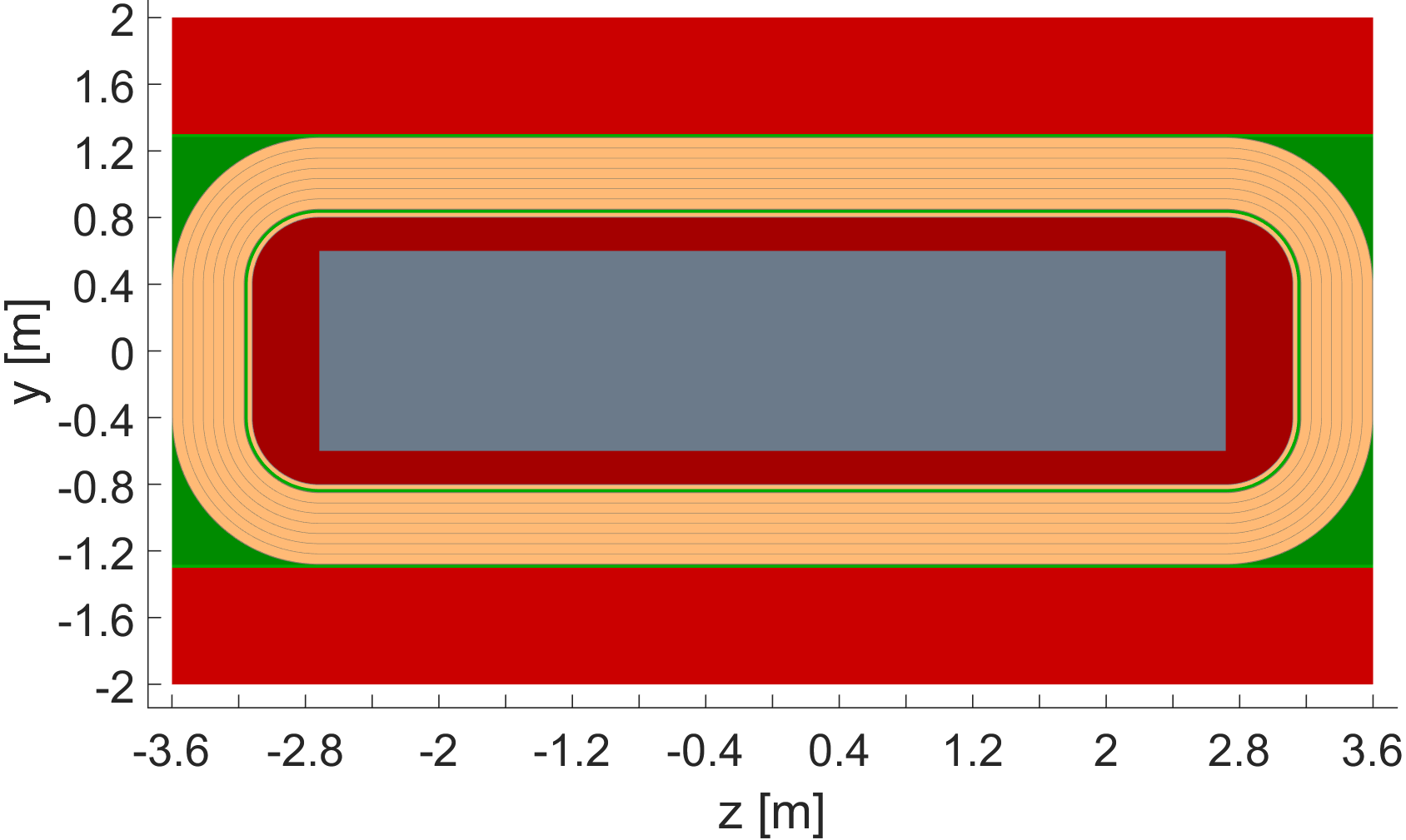} \raisebox{3.0cm}{{\small(a)}}}\\[1ex]
  \multicolumn2c{\rule{4ex}{0pt}\includegraphics[height=8.0cm]{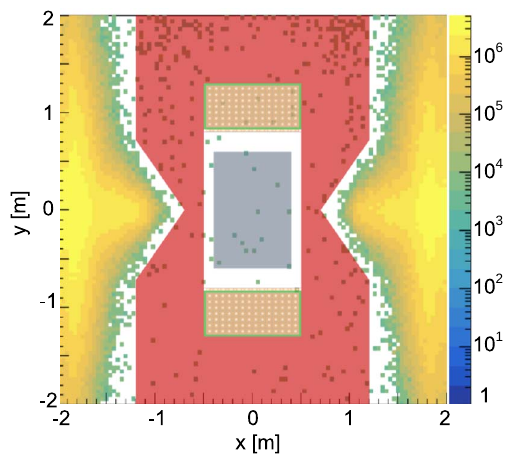}  \raisebox{4.2cm}{{\;\small(b)}}}\\[1ex]
  \includegraphics[width=0.58\linewidth]{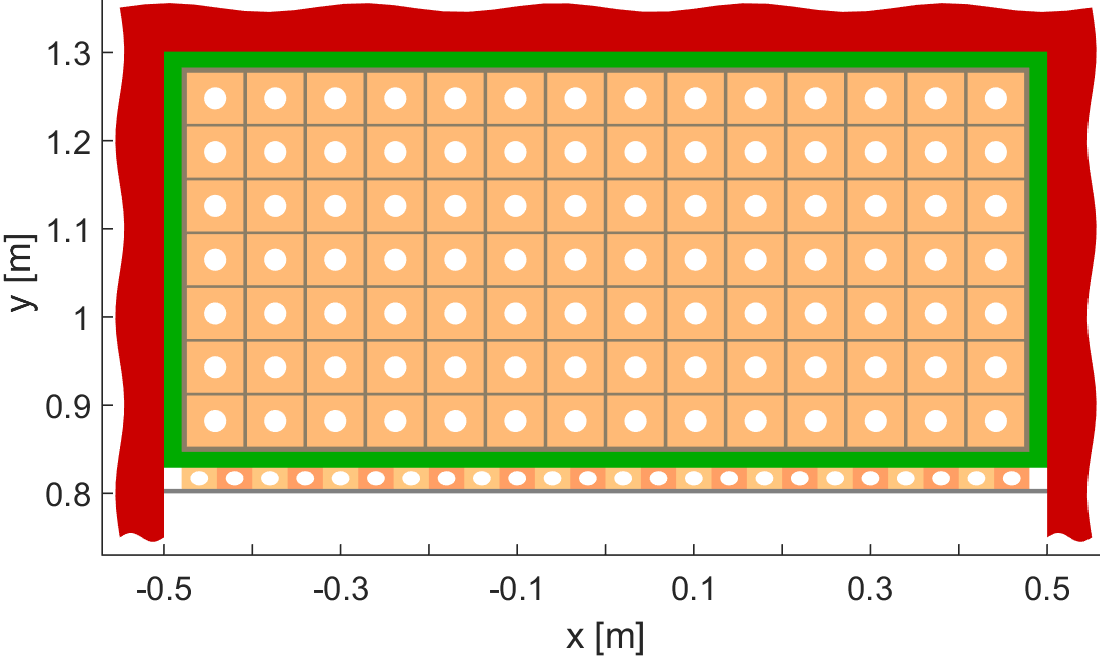} \raisebox{7.9em}{{\;\small(c)}}
    & \raisebox{4.2em}{\includegraphics[width=0.24\linewidth]{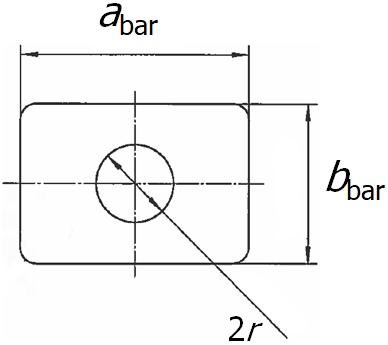}} \raisebox{7.9em}{{\;\small(d)}}
\end{tabular}
\caption{\label{fig:coilzy:xy:magnxymuon}
  Cross-sections of the SND magnet:
  (a) in the plane $x=0$;
  (b) in the plane $z=0$, superimposed to the simulated muons flux distribution
    in the transverse plane immediately downstream of the last sweeper magnet~\cite{shipc2019};
  (c) cut-out of the upper part in the plane $z=0$;
  (d) particular of the hollow bar type used as coil conductor.
  The gray box in (a--b) represents the instrumented region,
  where the mean flux density is specified as $B=1.25$\;T.
  }
\end{figure}

After the yoke has been determined in its size, shape and material we can afford
the detailed design of the coil. It has to comply with the following additional constraints or criteria:
\begin{enumerate}
\item \emph{Coil cross-section.} The total height of the magnetized volume
  $b = 1.60$ m is specified (table \ref{tab:magnet_spec_0}). Therefore, the
  total thickness is $t =  y_\text{max} -b/2 -h = 0.5$ m. However, the gross area $at$
  is not fully available to the coil (see figure \ref{fig:coilzy:xy:magnxymuon}). The coil thickness $t_\text{coil}$ is less than $t$ to accommodate thermal shield, insulating laminates, mechanical supporting laminate in about 8 cm (figure \ref{fig:coilzy:xy:magnxymuon}c). Similarly, the coil width $a_\text{coil}$ is less than $a$, so as to leave about 4 cm of lateral space for the tie-rods that fix the coil to the iron and for thermal insulation ~\cite{sgobba2011}.
\item \emph{Winding type.} A continuous double pancake coil configuration is assumed,
  so that all electrical and water pipe junctions are  external to the magnet.
\item \emph{Voltage.} The electric voltage $V$ at coil terminals should be as
  close as possible to 100 V so as to exploit synergy for power converters used at CERN.
\item \emph{Current.} The electrical current $I$ should be less than 14.4 kA (so as to have no more than two standard 8 kA converter modules with a 10 \% margin for control).
\item \emph{Cooling water temperature.} The inlet temperature
  $T_\text{in} = (29 \pm 1)$\textcelsius\ is specified by the CERN EN-CV-INJ Department,
  whereas the outlet temperature $T_\text{out}$ should not exceed 60\textcelsius\ to avoid damage to the resin.
\item \emph{Cooling water speed.} To avoid erosion, corrosion and impingements, the speed $w$ should not exceed 3 m/s.
\item \emph{Reynolds number.} To get a
  moderately turbulent flow, the condition $2000 <\text{\emph{Re}}<10^5$ should be satisfied.
\item \emph{Pressure drop.} In the water circuit, the pressure drop $\Delta p$ should not exceed the limit of 10\:bar.
\end{enumerate}
%
%

The magnetomotive force $\mathcal F = NI$ of about 1\:MA has been
estimated in section~\ref{sec:an:mod} since, for the magnetic structure
and materials considered, it mainly depends on the desired field $B$ and the horizontal gap size $a$. 
The effective cross-section of the coils and the value of the magnetomotive
force are almost fixed by the above considerations. Therefore, as stated
in section~\ref{sec:an:mod}, the aluminum option is discarded in order to
minimize the ohmic power $P$, which for a copper coil and a realistic filling
factor $f=0.65$ is about 1\:MW. The requirement of a
total coil voltage $V$ of about 100 V leads to a current $I = P/V$ of about 10 kA, hence to a number of turns $N = \mathcal F/I$ of about 100.
The opportunity to have cooling pipes of circular cross-section
(hence coil turns of nearly square cross-section) and the ratio
between
$a_\text{coil}$ and $t_\text{coil}$,
which is about 2, lead to select $N_a=14$ pancakes with $N_b=7$ turns each.
That yields to $N = N_aN_b = 98 \approx 100$.
It is worth noticing that:
\begin{itemize}
\item $N_a=14$, an even number, is compatible with the double pancake configuration;
\item greater values of the number of turns, e.g. $N_a=16$ with $N_b=8$, would make the design of the cooling system more cumbersome and increase V above 100 V;
\item the lower value of the number of turns $N=72$, with $N_a=12$ and $N_b=6$,
  still compatible with the constraint $I<14.4$ kA, would unnecessarily reduce
  the voltage well below 100\:V, while increasing the cross-section of the single
  turns, which might yield problems when bending the conductor.
\end{itemize}

The next step is to specify the cross-section of the hollow bars, followed
by the design of the cooling system. This design started from
a first guess of the parameters
and it went through a few iterations exploiting  the results of more accurate numerical electrical and thermal
analyses, as it will be shown in the next section.
The selected design configuration is then reported in table~\ref{tab:coil:confs}
in comparison with the LHCb magnet. The 3-D analyses reported in the next section show that all constraints are satisfied. 
As expected, the value of the ohmic power $P=1.02$\:MW is not far from the figure
provided by the procedure based on lumped parameters. However, it is worth noticing
that the ohmic power might further be optimized. Indeed, a significant reduction
(about 10\:\%) can be obtained by relaxing the stray field limit to 15\:mT,
while selecting a different magnetic material and a variable thickness of the yoke (different values of top and side thickness).

%
\begin{table} 
\centering
\begin{small}
\renewcommand{\tabcolsep}{0.9ex}
\caption{\label{tab:coil:confs}%
  Reference design configuration of the detector and comparison with the main parameters of the LHCb magnet.
  }
\begin{tabular}{l@{\,}lll c   cc c@{}}
  \toprule
  \mltclmn4{@{}r@{}}{ }                                                                           &&        SND  && LHCb~\cite{andre2002,andre2004,losasso2006}\\
  \cmidrule(l{0pt}r{.0em}){1-4} \cmidrule{6-6} \cmidrule{8-8} 
  \mltclmn4{@{}l@{}}{General magnet properties}\\
    &total power                    & $P$                                            & [MW]       &&       1.03  && 4.2\\
    &magnet efficiency              & $\eta$                                         & [-]        &&      .981   \\
    &internal {\blue usable} space along $z$ & $c_i$                                          & [m]        &&      5.43   \\
    &yoke thickness                 & $h$                                            & [cm]       &&     70      \\
    &max top stray field            & $B_\text{stray\,max}$                          & [mT]       &&     10    \\
    &max side stray field$^\dag$  &                                                & [mT]       &&     9    \\
    &total iron mass                &                                                & [t]       &&    356    && $\approx 1500$\\
  \mltclmn4{@{}l@{}}{\rule{0pt}{1.5em}Coil}\\
    &hollow bar material            &                                                &            &&          Cu && Al-99.7\\
    &n.\ of pancakes                & $N_a$                                          & [-]        &&          14 &&$2\cdot15$\\
    &turns per pancake              & $N_b$                                          & [-]        &&           7 && 15\\
    &total turns                    & $N=N_aN_b$                                     & [-]        &&          98 &&$2\cdot225$\\
    &hollow bar width               & $a_\text{bar}$                                 & [mm]       &&          64 && 50\\
    &hollow bar height              & $b_\text{bar}$                                 & [mm]       &&          58 && 50\\
    &water hole diameter            & $2 r$                                          & [mm]       &&        25.5 && 25\\
    &average turns length           & $l_t$                                          & [m]        &&        16.6 && 19.3\\
    &total winding length           & $l_\text{tot}$                                 & [km]       &&         1.6 && 8.7\\
    &total hollow bar mass          & $m_\text{tot}$                                 & [t]        &&          46 && $\approx 2\cdot25$\\
    &coil thickness                 & $t_\text{coil}$                                & [cm]       &&        43.6 \\
    &total thickness                & $t$                                            & [cm]       &&        50.1 \\
    &insulator/holes ratio          & $S_\text{i}/(N\pi r^2)$                        & [-]        &&        1.12 \\
    &coil fill factor               &\blue$f_\text{coil}=S_\text{active}/(a_\text{coil}t_\text{coil})$& [-]        &&         .75 \\
    &total fill factor              &\blue$f=S_{\text{active}}/(at)$                   & [-]        &&         .62 \\
  \mltclmn4{@{}l@{}}{\rule{0pt}{1.5em}Electrical and magnetic properties}\\
    &{magnetomotive force}          & $\mathcal F=NI$                                & [MA]       &&        1.014&& $2\cdot1.3$\\
    &current per turn               & $I$                                            & [kA]       &&        10.3 &&5.85 (6.6 max)\\
    &voltage                        & $V$                                            & [V]        &&        99   && 730\\
    &current density                & $J$                                            & [A/mm$^2$] &&        3.2  && 2.9\\
    &total resistance               & $R$                                            & [m$\Omega$]&& 9.6 @ 42.5\:$^\circ$C && 130 @ 20\:$^\circ$C\\
    &inductance                     & $L$                                            & [H]        &&       0.18  && 1.3\\
    &stored magnetic energy         & $W_m$                                          & [MJ]       &&       9.7   && 32\\
  \mltclmn4{@{}l@{}}{\rule{0pt}{1.5em}Double pancake configuration and cooling}\\
    &continuous bar length          & $l_\text{wc}=2N_bl_t$                          & [m]        &&        233  && 290$^\ddag$\\
    &parallel water circuits        & $N_\text{wc}=N_a/2$                            & [-]        &&          7 \\
    &inlet-outlet temperature raise & $\Delta T$                                     & [$^\circ$C]&&         25  && 25\\
    &total cooling flow             & $q_\text{tot}$                                 & [m$^3$/h]  &&         35  && 150\\
    &water speed                    & $w$                                            & [m/s]      &&        2.7  \\
    &Reynolds number                & \emph{Re}/1000                                 & [-]        &&        98   \\
    &pressure drop                  & $\Delta p$                                     & [bar]      &&        6.8  && 11\\
  \bottomrule
  \mltclmn{8}l{${}^\dag$ attained at $x=a/2$, $y=1.1$\:m (see figure~\ref{fig:3D:abcd}c)}\\
  \mltclmn{8}l{${}^\ddag$ the LHCb magnet has a single pancake configuration}
\end{tabular}
\end{small}
\end{table}
%

Finally in table~\ref{tab:models_comparison}
we compare main design figures
calculated with the analytical model of section~\ref{sec:an:mod} and the accurate numerical model.
Such comparison assumes iron ST 1010 ATLAS choice,
a mean turns length $l_t=16.6$\:m and a filling factor $f=0.62$
as accurately determined with the numerical model and reported in table~\ref{tab:coil:confs}.
A very good agreement can be recognized.

\begin{table}[tbp]
\centering
\caption{\label{tab:models_comparison}Comparison of the main design parameters from different modeling approaches.}
\begin{tabular} {@{}llcc@{}}
\toprule
                        &      & Analytical   & FEM 3-D\\
\midrule
$B$                     & [T]  & 1.25  & 1.25\\
$NI$                    & [MA] &1.01 & 1.014\\
${B}_\text{Fe}$         & [T]  & 1.75   & 1.73\\
${B}_\text{stray\,max}$ & [mT] & 11    & 10\\
\bottomrule
\end{tabular}
\end{table}
%

\subsubsection{Thermal shield}
\begin{figure}[tbp]
\centering 
\includegraphics[height=5cm]{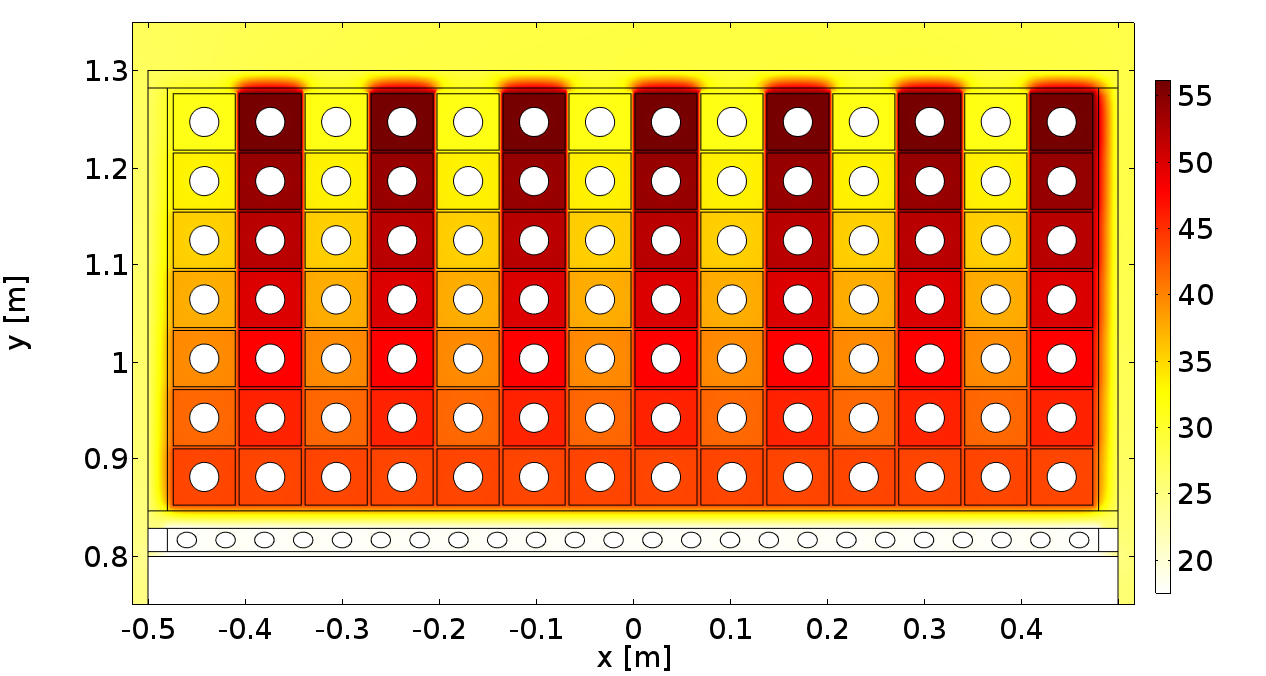}
  \rule{-0.89cm}{0pt}\raisebox{4.60cm}{{\footnotesize[$^\circ$C]}} \rule{+0.7cm}{0pt}
\caption{\label{fig:therm:2D}
  2-D thermal numerical simulation for the reference configuration (table~\ref{tab:coil:confs}).
  The maximum temperature raise,
  $\Delta T=25^\circ$C, occurs between the outermost turns of odd and even pancakes.
  A differential thermal expansion of about 3\:mm is calculated along the major magnet length $c$.
  }
\end{figure}
Figure~\ref{fig:coilzy:xy:magnxymuon}c
shows that the coil is thermally insulated.
The proposed insulator is Vulkollan${}^\text{\textregistered}$\ or a similar product,
which has excellent mechanical properties, including elastic
ones, to accommodate the
{different thermal expansion of the yoke}.
The inner and outer insulator thickness
shown in green 
is taken as 18\,mm, while the side one is 20\,mm.
The insulation layer plays also the role of reducing the temperature of the yoke,
preventing magnetic ageing issues~\cite{sgobba2011}.

The additional single-layer copper circuit shown in the lower part of
figure~\ref{fig:coilzy:xy:magnxymuon}c,
in contact
with the inner coil insulation and with a supporting 5\,mm thick non-magnetic
steel laminate, is a thermal shield, hence not fed with any electric current,
used to insulate the instrumented region,
and keeps it at about $18\,^\circ$C.
Such shield is made of copper hollow bars with rectangular cross-section
with the following characteristics:
two continuous even/odd parallel water circuits, each made of
12 turns and 181\,m long, with corresponding
inlet water pipes connected at opposite sides ($x\approx\pm0.5$\,m),
to achieve a uniform temperature;
total mass 2.3\,t;
input thermal power (from coil) about 6\,kW;
inlet/outlet water temperature 17/19\,$^\circ$C;
cross-section area $40\times24$\,mm$^2$;
elliptic cooling hole with major/minor diameter equal to 20/16\,mm$^2$
and hydraulic diameter $d_h=17.7$\,mm;
water speed 1.56\,m/s;
Reynolds number about 38000;
pressure drop 3.1\,bar.
The hydraulic diameter is given by four times the area divided by the perimeter
of the (wetted) pipe cross-section.
In the case of elliptic cross-section the perimeter can easily be calculated by means of
standard special
functions~\cite{nist2010}.

Figure~\ref{fig:therm:2D}
reports the result of a 2-D thermal numerical simulation for the design option of table~\ref{tab:coil:confs}.
The double pancake configuration implies that the maximum temperature difference,
$\Delta T=25\,^\circ$C, is attained between the outermost turns of odd and even pancakes.
The consequent differential thermal expansion along the major magnet length, $c$, is about 3\;mm.
The resin encapsulating the coil will have to withstand such differential expansion.

\subsection{3-D Field maps\label{sec:elmag:FEM}}
\begin{figure}[tbp]
\centering 
\includegraphics[height=4.8cm]{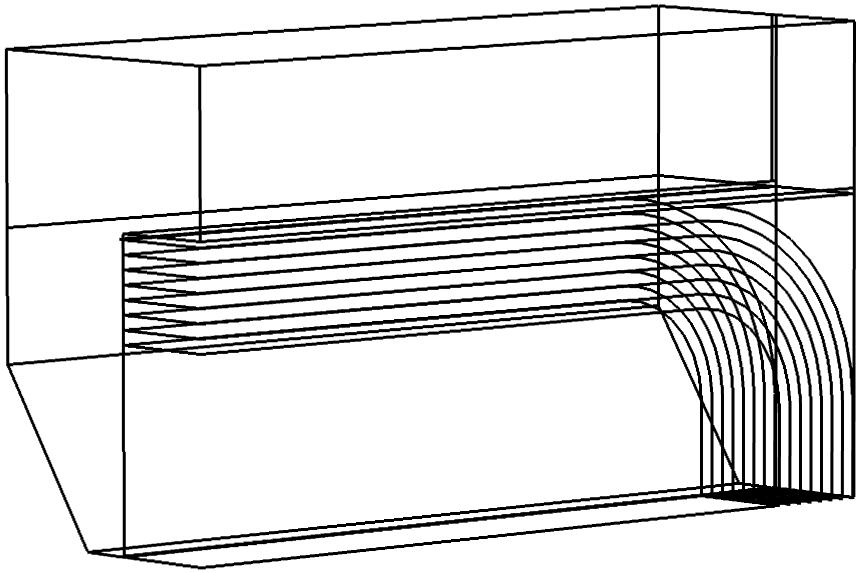}
\caption{The 3-D model of the reference design }
\label{fig:3_D_COMSOL_model}
\end{figure}

We report here the
results of a detailed 3-D
simulation of the electromagnetic problem,
after the definition of the reference design as described in previous sections.
Sizes and specifications are reported in table~\ref{tab:coil:confs}.
In figure~\ref{fig:3_D_COMSOL_model} the structure of the FEM model is sketched,
with core and coil details.
The magnetic curve ST 1010 ATLAS fit shown in figure~\ref{fig:different:iron} is assumed as reference iron model.
Due to the symmetry only one eight of the entire structure is simulated,
hereafter named block; on the corresponding cut boundaries the symmetry condition is imposed,
as well as the magnetic insulation at the external region boundaries.
Such block is meshed with a total of about 239000 nodes, of which about 102000
for the air gap region, 55000 for the iron yoke, 12600 for the coil and the remaining for the external region.
\begin{figure}[tbp]
\centering 
\includegraphics[height=4.8cm]{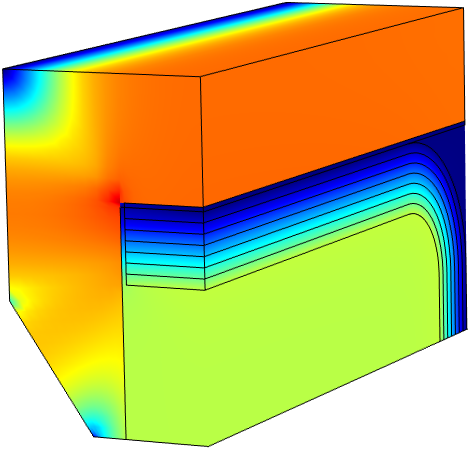}
  \rule{1cm}{0pt} \includegraphics[height=5.0cm]{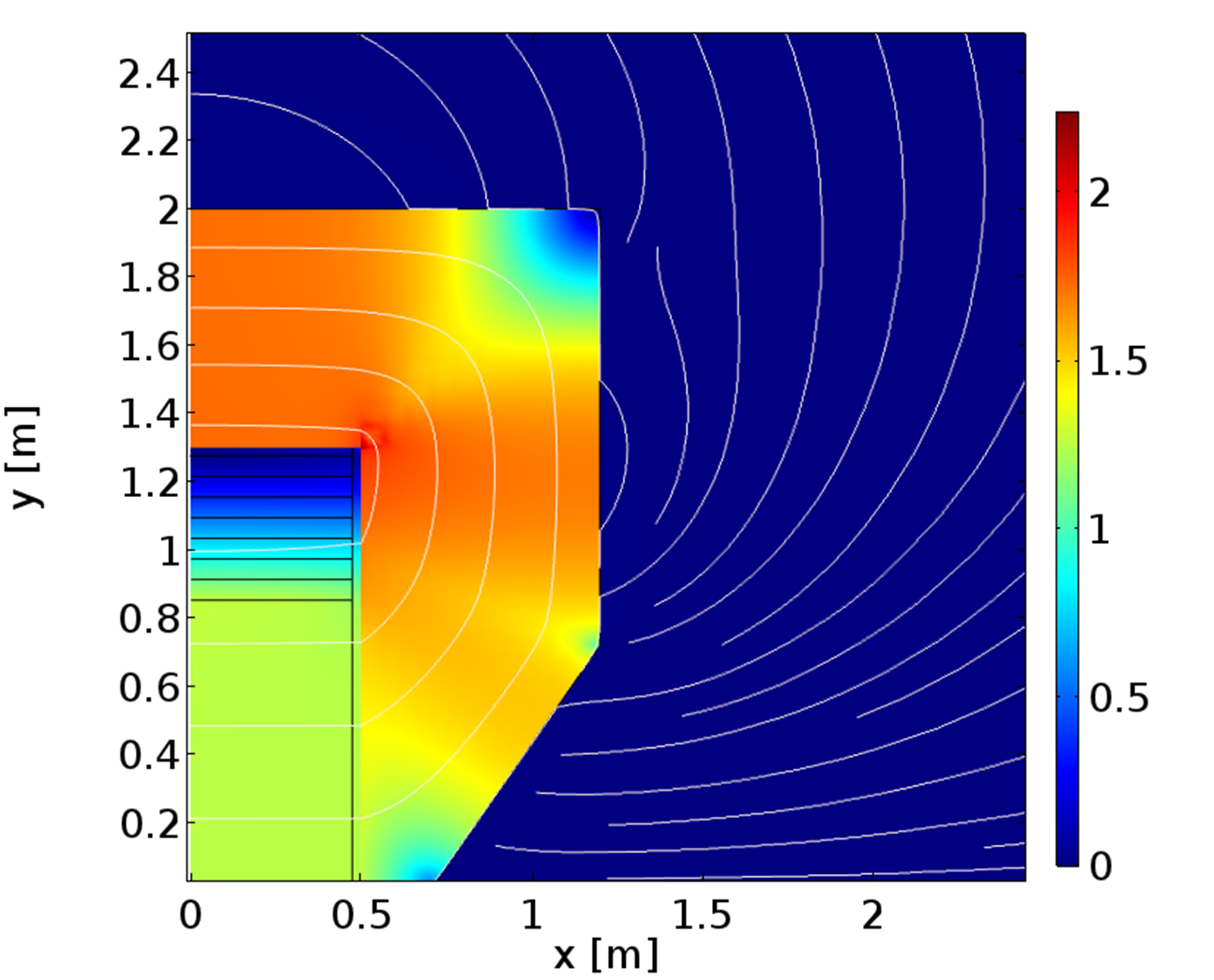} \rule{-0.9cm}{0pt}\raisebox{4.7cm}{{\small[T]}}
\caption{$\lvert B\rvert$ mapping within the magnet and outside: 3-D view (left), 2-D section at $z=0$ (right).
  The point $(0,0,0)$ is the center of the magnet.}
\label{fig:3_D_COMSOL_B_field_internal}
\end{figure}
\begin{figure}[tbp]
\centering 
\begin{tabular}{@{}r@{\;\;\;}l@{}}
  \includegraphics[width=7.5cm]{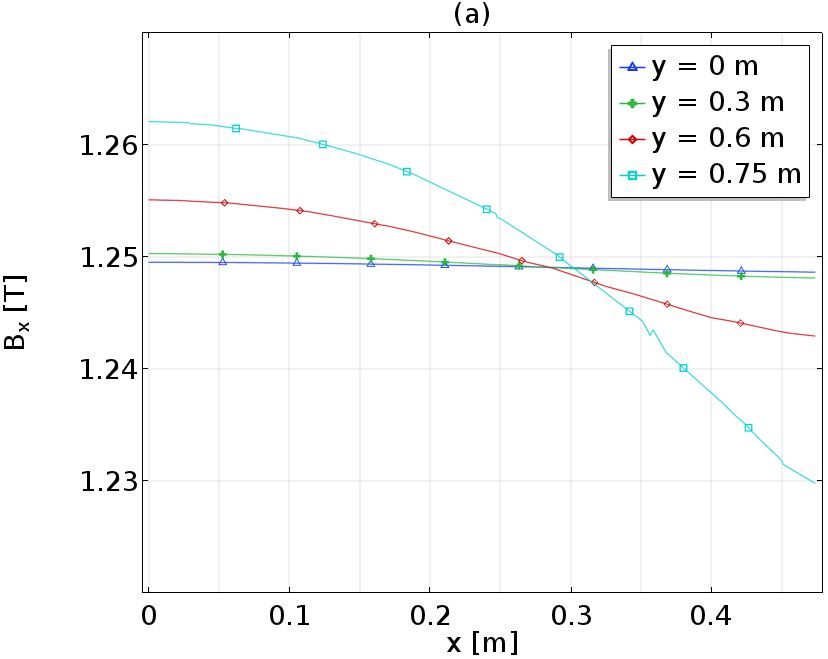}
    & \raisebox{0.25cm}{\includegraphics[width=7.20cm]{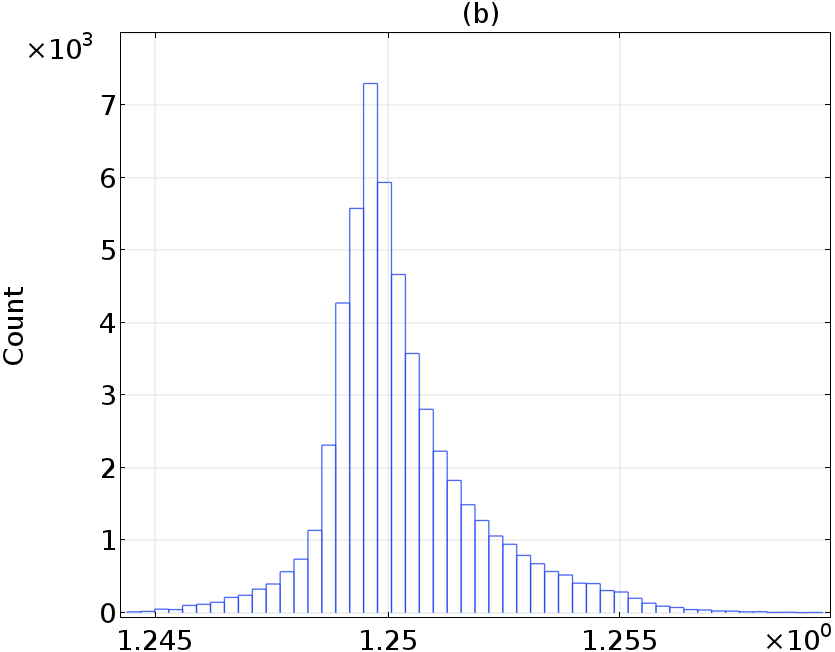}}
      \raisebox{0.01cm}{\rule{-3.25cm}{0pt}{\scriptsize[T]}}\\[0.2cm]
  \includegraphics[width=7.30cm]{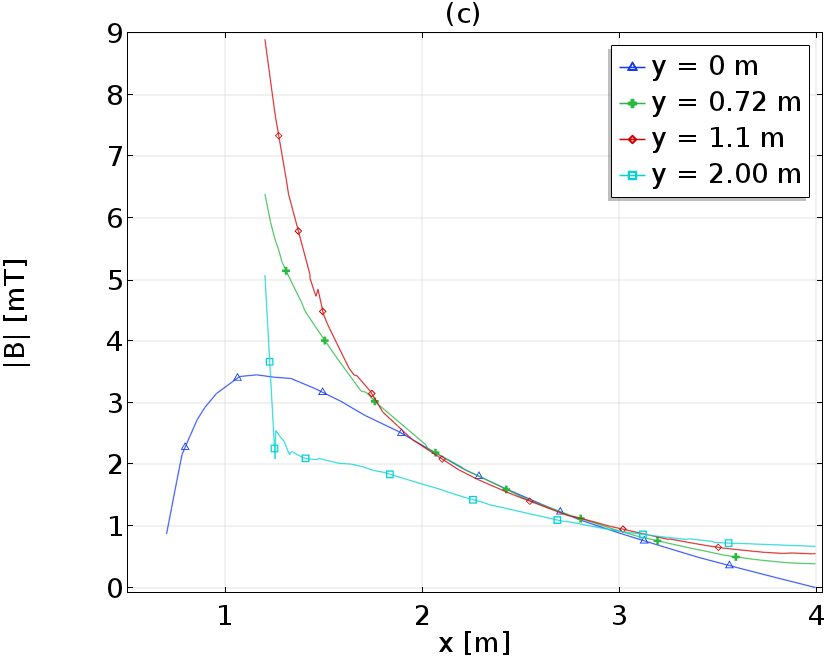}
    & \rule{-0.15cm}{0pt}
      \raisebox{-0.01cm}{\includegraphics[width=6.93cm]{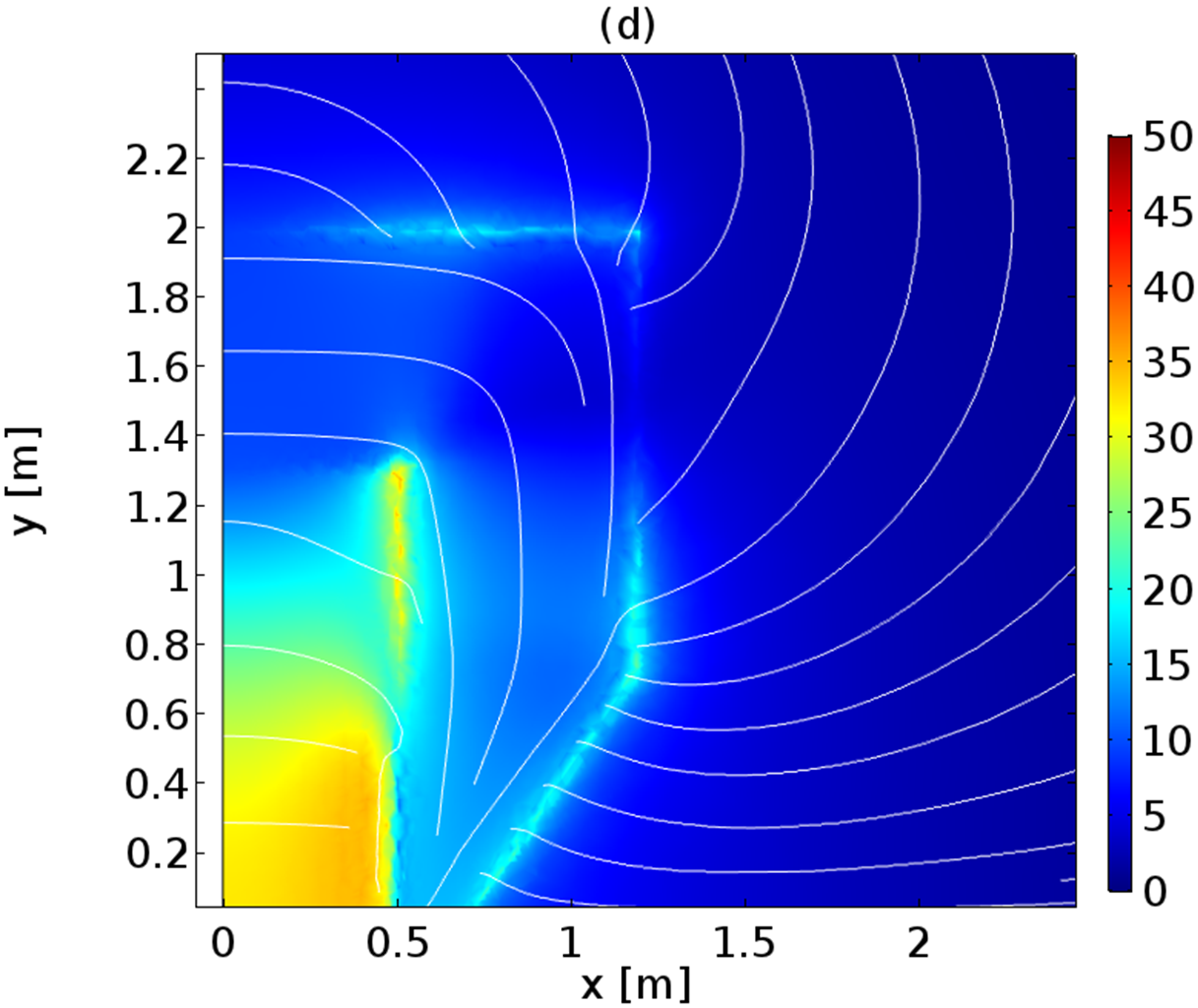}}
      \raisebox{5.4cm}{\rule{-0.55cm}{0pt}{\scriptsize[mT]}}
\end{tabular}
\caption{\label{fig:3D:abcd}
  (a) $B_x$ as a function of the horizontal axis $x$ at the $z=0$ section,
  for different horizontal lines at $y=0$, $y=0.3$, \mbox{$y=0.6$}, $y=0.75$\:m.
  {\blue (b) Distribution of the $|B|$ values on the mesh points in the detector region.}
  (c) $\lvert B_{\text{stray}}\rvert$ as a function of $x$ at the $z=0$ section,
      for different horizontal lines at $y=0$, $y=0.72$, $y=1.1$, $y=2.00$\:m.
  (d) Stray field map $\sqrt{\smash[b]{B_{\text{stray}x}^2+B_{\text{stray}y}^2}}$
      at the sections $z=\pm(c/2+2\:\text{cm})$, namely 2\:cm upstream/downstream of the magnet.
  }
\end{figure}
\begin{table}[tbp]
\centering
\caption{\label{tab:3D:Bstray:surf}{
  3-D FEM analysis.
  {\blue Stray field}
  at various $z=\text{constant}$
  cross-sections and relevant yoke surface locations.
  $z=0$ and $z=2.7$\:m correspond to the center and the end of the detector region
  (the gray box in figure~\ref{fig:coilzy:xy:magnxymuon}a).}}
\begin{tabular} {@{}l@{\;\;}l@{} cc @{}l @{~}lccc@{}}
\toprule
\multicolumn2{@{}l}{$xy$ yoke surface location} & &\quad& $z$                   & [m]  & 0 & 1.35 & 2.7\\ 
\cmidrule{1-2} \cmidrule{5-9}
$@$ top    &($x=0$, $y=y_\text{max}$)      & &     & $B_\text{stray\,max}$ & [mT] & 10 & 10 & 10\\
$@$ side   &($x=a/2$, $y=1.10$\:m)          & &     & $B_\text{stray}$ & [mT] &  9 &  8 & 7.5\\
$@$ septum &($x=1.00$\:m, $y=0.43$\:m)      & &     & $B_\text{stray}$ & [mT] &  5 &  5 & 4.5\\
$@$ max muons flux &($x=1.80$\:m, $y=0$)       & &     & $B_\text{stray}$      & [mT] &  3 &  3 & 3\\
\bottomrule
\end{tabular}
\end{table}

The FEM simulation for the set of used parameters is reported in
figure~\ref{fig:3_D_COMSOL_B_field_internal}, where the modulus of flux density $B$ is given in a 3-D view and 2-D section, respectively. The complete
mapping of the field allows to evaluate the field uniformity within
the detector region, and provide some local information at specified section/lines.
Figure~\ref{fig:3D:abcd}a
reports the value of $B_x$ at the $z=0$ section for different horizontal lines.
Figures~\ref{fig:3D:abcd}\,(a-b)
show that the field uniformity $|\Delta B/B|$ in the internal region
is limited to $\pm1$\:\%, in agreement with the requirements.
%
Figure~\ref{fig:3D:abcd}c
shows the external field $B_\text{stray}$ as a function of $x$
at the $z=0$ section for different horizontal lines,
starting 1\:mm away from the yoke.
The line at $y=0$ starts at lower $x$ values because of the septum shape. 
The limit { $B_\text{stray\,max}<10$\:mT} is fully accomplished as expected.
Finally, in figure~\ref{fig:3D:abcd}d
a
suggestive
picture of the
$\sqrt{\smash[b]{B_{\text{stray}x}^2+B_{\text{stray}y}^2}}$
field map is given at a vertical section immediately
upstream of the magnet,  2\:cm outside. 
{\blue Table~\ref{tab:3D:Bstray:surf}
reports stray field values at relevant yoke surface locations.}


\section{Mechanical issues\label{sec:mech:issues}}

\subsection{Forces and stresses analysis\label{sec:force:stress}}
\begin{figure}[tb]
\centering
\includegraphics[scale=1.2]{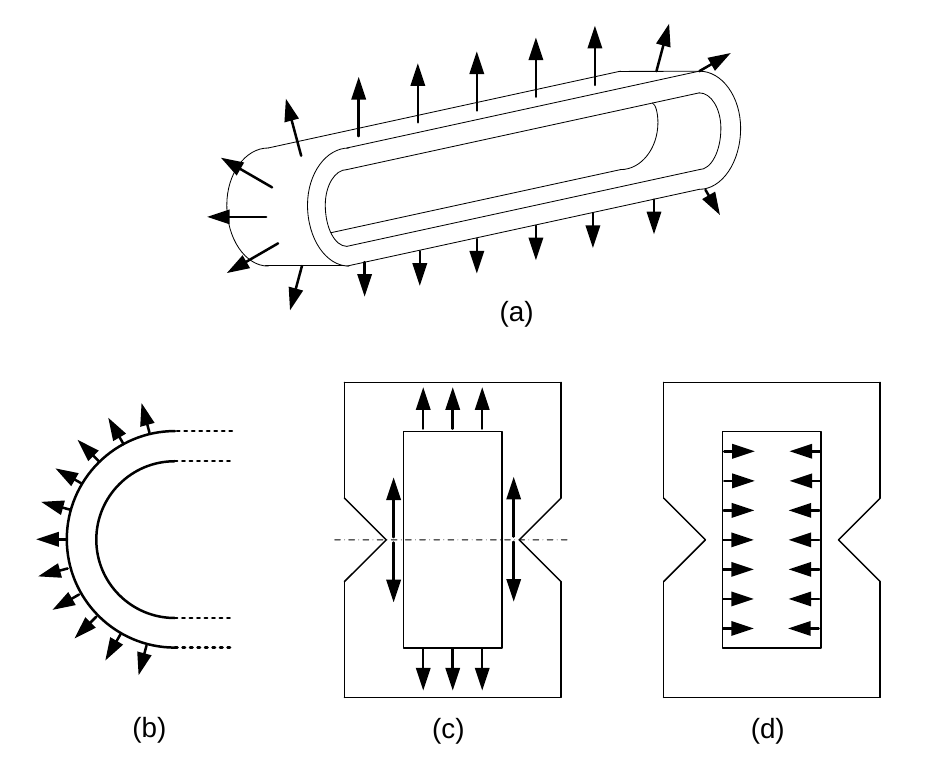}
\caption{\label{fig:Forces and stresses}
a) Magnetic coil self-force. b) Self-force stress at the coil edges. c) Iron stress induced by the coil vertical force. d) Horizontal magnetic force on the iron. }
\end{figure}

In order to complete the design, we have to consider the
problem of the magnetic force and the corresponding induced stresses~\cite{montgomery1969},
due to
Lorentz force on the coil that tend to burst the coil
radially outward and crush it axially.
In figure~\ref{fig:Forces and stresses}
a visual representation of such effects is given.

A fair evaluation of the total force can be obtained as that 
produced by an infinitely thin current sheet carrying
the total current. In this way, following Maxwell's stress tensor method~\cite{Stratton1941},
the magnetic force can easily be calculated by means of the magnetic pressure at the internal coil boundary as:
\begin{equation}
p_{\text{mag}} = \frac{B^2}{2\mu_0}. 
\label{eq:magnetic_pressure}
\end{equation}
where ${B}$ is the reference induction field within the chamber volume.
{\blue
This expression remains valid for the case of thick conductors,
for which it
can be thought as the difference at the inner and outer edges of the coil.} 
%
\tblue{Equation~\eqref{eq:magnetic_pressure} consents}
to calculate the stress on the coil bent section,
as well as the stresses on the straight sections transmitted to the iron,
without dealing with the distributed body forces.
Also the horizontal force pulling the vertical iron arms inward, and the corresponding
induced stress, can be directly estimated by means of the magnetic pressure concept.

\subsubsection{Analytical models\label{sec:0D:force:stress}}
The evaluation of the coil stress at the bent edges is done by treating
the coil as a thick-walled cylinder supporting the corresponding
internal pressure of a gas.
Using
Lamé equations~\cite{Timoshenko1956},
which give the stresses for thick-walled cylinders as a function of the radius $r$,
and neglecting the external air pressure compared to the internal one,
the tangential stress  $\sigma_t$ is expressed as:
\begin{equation}
\sigma _t  = p_\text{mag}\frac{r_i^2 }{r_e^2 -r_i^2}\biggl(1+\frac{r_e^2}{r^2}\biggr)
\end{equation}
where $r_i=0.8$\:m and $r_e=1.3$\:m are the inner and outer coil radii, respectively.

The maximum tangential stress, that is the greatest magnitude of direct stress,
amounts to 1\:MPa, therefore well below the yield strength of the copper of about
50\:MPa at $40^\circ$C.
It has to be remarked that the real profile of the coil edge will slightly differ
from the semi-cylindrical one in order to increase the inner volume available
for the detectors. Nevertheless, the corresponding stresses are not expected to
vary significantly. A more detailed analysis will be presented in
section~\ref{sec:stress:3D}.

To evaluate the vertical force transmitted to the iron by the horizontal sections of
the coil, the internal magnetic pressure has to be multiplied by the proper surface.
The total resulting force on the upper part of the yoke will be the magnetic one
reduced by the weights of the upper horizontal sections of both iron yoke and coil.
Such force, equally distributed between the two vertical arms of the yoke,
produces a maximum stress of about 1.5\:MPa at the minimal iron thickness
in the septum, that is well below the yield strength $\sigma_\text{y}=300$\:MPa,
which is the typical yield strength for standard iron.

The horizontal force pulling inward the vertical iron arms, and the corresponding
induced stress, are also calculated via Maxwell's stress tensor method. In this
case the magnetic pressure~\eqref{eq:magnetic_pressure} is pulling the vertical
inner yoke surface because the magnetic field is nearly perpendicular in the air side.
Then the bending moment is evaluated by assuming the vertical iron arm as a simply
supported plate under bending where one dimensional model can be used,  due to the typical ratio between the longitudinal $z$ and transversal $y$ dimension.
The bending moment is then calculated with respect to a supported beam subject
to the distributed horizontal load given by the magnetic pressure.
Using the flexure 
formula, under the
conservative assumption that the beam thickness coincides
with the minimal section at the yoke septum, the maximum stress results in about
20\:MPa, more than one order of magnitude below $\sigma_\text{y}$.


\subsubsection{3-D analysis\label{sec:stress:3D}}
The main mechanical stress on the structures is here analysed with 3-D FEM simulations.
Sizes and specifications are reported in table~\ref{tab:coil:confs}.

Coupled magnetic-structural finite element 3-D analysis allows a more
detailed assessment of the stresses due to the electromagnetic forces acting on
both the coil and the iron yoke. In particular, the analysis
returns the forces as distributed body loads overcoming the simplification
of the magnetic pressure employed in the preliminary analysis.
The coil
has been modeled  as a ``racetrack'' neglecting all the insulating
layers whereas the iron yoke has been considered as a single piece.
Note that, due to the presence of the floor, the bottom horizontal surface of the iron has
been considered fixed along $y$. Therefore, for the mechanical case, the simulation cannot be restricted to 1/8 of the structure.

\begin{figure}[tbp]
\centering 
\includegraphics[height=5cm]{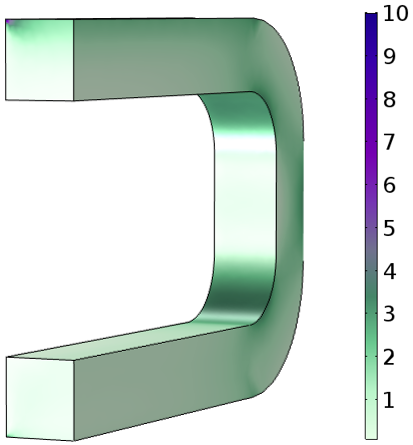}
  \rule{-0.0cm}{0pt}\raisebox{4.7cm}{{\small[MPa]}}
    \rule{2cm}{0pt}\includegraphics[height=5cm]{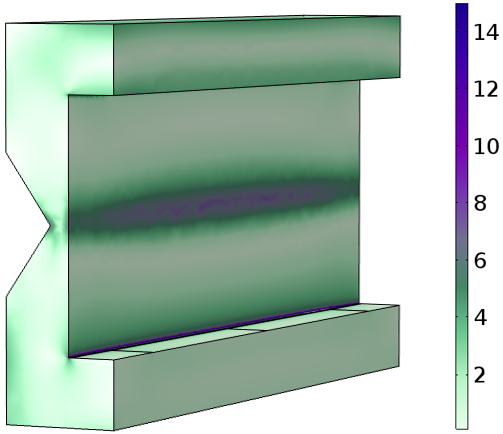}
      \rule{-0.0cm}{0pt}\raisebox{4.7cm}{{\small[MPa]}}\\
(a) \rule{7.5cm}{0pt} (b)
\caption{The 3-D model of copper (a) and iron (b) stress}
\label{fig:3_D_COMSOL_stress}
\end{figure}

Figure~\ref{fig:3_D_COMSOL_stress}a shows that the equivalent stress within the coil, evaluated according
to the von Mises criterion, $\sigma_\text{M}$, reaches a maximum value of about 3.4\,MPa in
relatively small regions of the bent end.
{\blue Compared to the analytic result,}
this is roughly a factor of 2 worse 
since the profile of the coil
edge slightly differs from the semi-cylindrical one through a straight
vertical section. 

Figure~\ref{fig:3_D_COMSOL_stress}b shows that the equivalent stress $\sigma_\text{M}$
within the iron yoke, reaches a value of about 8.3\:MPa
at the septum corresponding to the minimum iron section.
This value is about one third of that previously estimated with the conservative assumption of considering for the whole vertical iron arm the minimum iron thickness of the septum. 
%

\subsection{Some functional issues}
We finally discuss some additional issues that,
although not essential in the overall design as described above, are still relevant for more detailed design. It has to be recalled here that, beside the normal operation regime, the inner magnet volume
as described in section~\ref{sec:exp:req} requires to be accessed for the detector installation and maintenance. Some opening mechanism needs to be defined, allowing reliable, simple and fast operations.

\begin{figure}[htbp]
\centering 
\includegraphics[width=.9\linewidth]{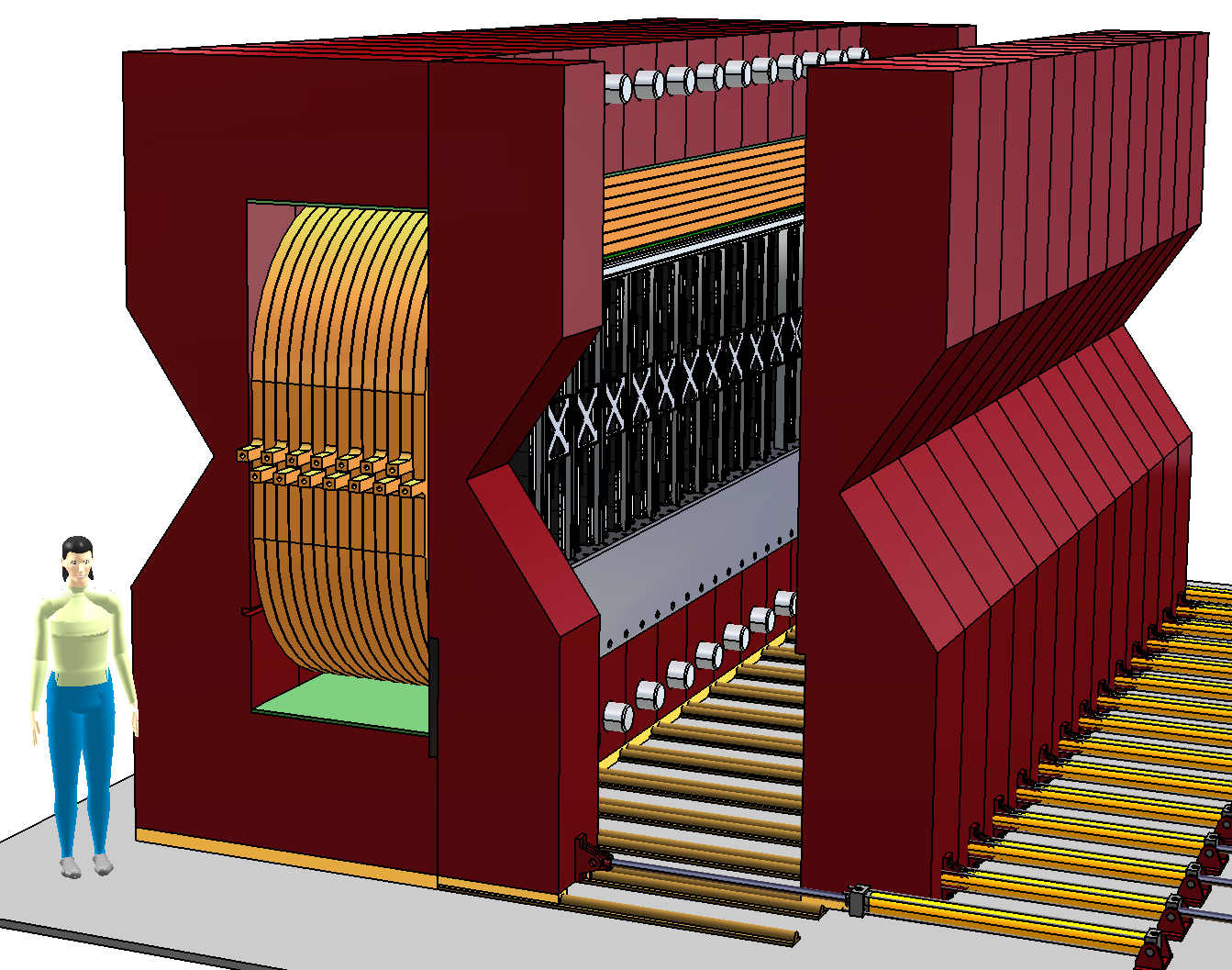}
\caption{Overall structure of the SND magnet, with partial view of the internal detectors.}
\label{fig:Magnet.pdf}
\end{figure}

Different schemes can be considered, that are compatible with the presented design. The significant amount of work needed for their detailed exploration and comparison largely exceed the scope of this paper. Nevertheless we would like to show some possible solution here, accomplishing the requirements, giving some insight to the related mechanical issues. Such proposed segmentation and opening scheme is 
depicted in figure~\ref{fig:Magnet.pdf},
where the iron yoke is split in independent parts, and a side opening is considered for each slab. The side slabs are coupled to the whole structure by means of dowels, and a undercarriage allows the lateral sliding.
We consider in the following the problem of sizing the coupling dowels, the opening force due to residual iron magnetization and the possible deformation of the structure when a prescribed number of slabs is removed.
%

\subsubsection{Structural dowels design\label{sec:dowels:dimensioning}}
We tackle here the dimensioning of the dowels connecting the vertical iron arms
to the upper and lower horizontal tracts of the yoke.
They have to resist to the shear stress induced by the vertical force
coming from the coil.  
The same assumptions (already considered in
section~\ref{sec:0D:force:stress})
that the magnetic pressure internal to the chamber produces the force
bursting the coil allow us to study the vertical force acting on the iron. Assuming
that the magnetic pressure is uniform within the chamber, the force
acting on the upper horizontal section of the iron yoke is evaluated
as the product of pressure and surface. The total force will be the difference
between the bursting force just calculated and the weights of the upper
horizontal sections of both iron yoke and coil.

In order to find the dowel section able to withstand the vertical
force acting on the iron we consider only the upper part of the iron yoke
modeling the horizontal section as an isostatic beam. Therefore, the mobile part
of the iron yoke has to balance the main force with a total constraint
reaction of about 1740\:kN.
This reaction has to be sustained by dowels of proper cross-section.
Assuming for the iron a $K_t$ (corrected yield strength) of 50\:N/mm$^2$
(``low strength'' iron), it is possible to find a total minimum surface
of $35\cdot10^3$\:mm$^2$ needed for the whole dowels.
For a 15 sections solution with the one dowel (see figure \ref{fig:Magnet.pdf}), 15 slabs, the diameter of the single dowel can be assumed to be 160\:mm (including safety factors).

\subsubsection{Opening force and deformations \label{sec:open:force}}
The force required to open the magnet when the current is turned off
(see section~\ref{sec:exp:req}) can approximately be calculated as follows.
Before opening the magnet, a current ramp down is performed, at the end of which
the $B$ field pattern can be assumed to be qualitatively the same as the one
corresponding to operation.
The condition $I=0$ implies
$aH+\ell H_\text{Fe}=NI=0$. Combining this equation with the flux
balance equation ~\eqref{eq:flux_bal},
while assuming $h\approx0.7$\:m, $t\approx0.5$\:m, $\ell\approx3a$,
provides $B_\text{Fe} =-\mu_0(c-t)(b+t)/(2hc)(\ell/a)H_\text{Fe} \approx-4\mu_0H_\text{Fe}$,
which is a line in the second quadrant of the plane $(H_\text{Fe},B_\text{Fe})$.
The worst-case condition is evaluated by assuming $H_\text{Fe}\approx H_\text{c}\approx200$\:A/m,
where $H_\text{c}$ is the coercive field.
That gives $B_\text{Fe}\approx1$\:mT, and in turn a force per unit surface
$B_\text{Fe}^2/(2\mu_0)\approx0.5$\:N/m$^2$, which is negligible.

Finally, as for reference, we calculated the worst case deformation of the structure when all the slabs are completely open, except for the two terminal ones, as shown in figure \ref{fig:Magnet.pdf}. The stress and deformation analyses for the open structure have been carried out with 3-D mechanical simulations, assuming an attachment boundary condition between the upper horizontal surface of the coil and the iron yoke. The maximum displacement for such case, attained at the top center of the structure on the opened face, is limited to about 30 $\mu m$, and the maximum Von Mises stress to about 6 MPa. Such values are fully compliant with admitted deformation of any involved structure and with the yield strength for both iron and copper.

The above analysis suggests no evident structural problem in the sectioning and opening scheme, at the considered detail level. The actual number of slabs as well as the opening scheme will be better specified and optimised in further design phases, according to specific requirements of the detector structure as well as to mechanical and manufacturing issues.

\section{Conclusions\label{sec:concl}}
A realistic design of the magnet for the SND detector of the
SHiP experiment, fully compliant with specifications and constraints has
been provided. Different options have been preliminarily considered,
defining the normal conducting copper solution as best suited to the
problem for different order of reasons, from structural ones to
resilience, reliability and maintenance.

Due to limitations in size and shape for the coil and yoke, the design
task, basically played between the conflicting goals of high magnetic
efficiency and minimal power consumption, revealed the need for some
deepening of standard analytical design tools. The design optimization
steps have been defined and described in detail, trying to give deep insight in the process. 

Such developments have been the guidance for the 3-D FEM analysis, that
has assessed the figures of merit and the general quality of the
established design option. In particular a detailed design set of design parameters
is given, fulfilling all the requirements and constraints.

Finally, beside the fundamental electromagnetic, thermal and mechanical
analysis, some basic manufacturing issues related to the required
accessibility of the SND along with realistic
solutions have been described.


\acknowledgments
The Authors wish to thank
Gilles~Le~Godec and Serge~Deleval
for their extensive support concerning
standards and best design practice, with reference to power converters
and cooling, respectively.
The Authors are grateful to
Attilio Milanese, Rosario Principe, Vittorio Parma, Pierre-Ange Giudici,
Vitalii Zhiltsov, Jakub Kurdej, Isabel Bejar Alonso and Marco Buzio 
for fruitful discussion and useful suggestions.
This work is supported by a Marie Sklodowska-Curie Innovative Training Network Fellowship
of the European Commissions Horizon 2020 Programme under contract number 765710 INSIGHTS.



\end{document}